\input{aipcheck}

\documentclass[
    ,final            % use final for the camera ready runs
  ]
  {aipproc}

\layoutstyle{8x11single}

\def\apj{ApJ.}
\def\aj{AJ.}
\def\nat{Nature.}

\begin{document}

\title{Investigating the Possibility of Screening High-z GRBs based on BAT Prompt Emission Properties}
%\title{Screening High-z GRBs with BAT Prompt Emission Properties}



\author{T. N. Ukwatta}{
  address={The George Washington University, Washington, D.C. 20052}
  ,altaddress={NASA Goddard Space Flight Center, Greenbelt, MD 20771}
}

\author{T. Sakamoto}{
  address={NASA Goddard Space Flight Center, Greenbelt, MD 20771}
  ,altaddress={The University of Maryland, Baltimore County, Baltimore, MD 21250}
%  ,altaddress={Oak Ridge Associated Universities, P.O. Box 117, Oak Ridge, Tennessee 37831-0117}
}

\author{K. S. Dhuga}{
  address={The George Washington University, Washington, D.C. 20052}
}

\author{W. C. Parke}{
  address={The George Washington University, Washington, D.C. 20052}
}

\author{S. D. Barthelmy}{
  address={NASA Goddard Space Flight Center, Greenbelt, MD 20771}
}

\author{N. Gehrels}{
  address={NASA Goddard Space Flight Center, Greenbelt, MD 20771}
}

\author{M. Stamatikos}{
  address={NASA Goddard Space Flight Center, Greenbelt, MD 20771}
  ,altaddress={Oak Ridge Associated Universities, P.O. Box 117, Oak Ridge, Tennessee 37831-0117}
}

\author{J. Tueller}{
  address={NASA Goddard Space Flight Center, Greenbelt, MD 20771}
}

%\author{K. S. Dhuga on behalf of the Swift-BAT team.}{
%  address={The George Washington University, Washington, D.C. 20052}
%}

\begin{abstract}
Being able to quickly select among gamma-ray bursts (GRBs) seen by
the Swift satellite those which are high-z candidates would give
ground-based observers a better chance to determine a redshift for
such distant GRBs.  Information about these high-z GRBs is
important in helping to resolve questions about the early universe
such as the formation rate of high-z GRBs, the re-ionization
period of the universe, the metallicity of the early universe, and
the Hubble expansion. Initially using a sample of 51 GRBs with
previously measured redshifts, we have developed high-z screening
criteria employing the GRB spectral as well as temporal
characteristics of the prompt emission from the Burst Alert
Telescope (BAT) on Swift.
%With our selection, we were able to
%issue high-z alerts to the GRB community based on these criteria.
Now that the sample has increased to 81 GRBs, we have revisited
the screening criteria and our methodology. Our updated high-z
screening criteria are presented in this paper.
\end{abstract}

\maketitle

%%%%%%%%%%%%%%%%%%%%%%%%%%%%%%%%%%%%%%%%%%%%
%% MAINMATTER
%%%%%%%%%%%%%%%%%%%%%%%%%%%%%%%%%%%%%%%%%%%%

\section{Introduction}
The detection of high redshift ($z$) Gamma-Ray Burst (GRBs)
promises to give us valuable information about the early universe.
Both GRB prompt emission and afterglows are so powerful that they
should be detectable out to redshift of $z$ $>$ 10
\citep{2000ApJ...536....1L}. Being the brightest explosions yet
seen in the universe since the Big Bang, GRBs are well-suited
objects to probe the evolution of cosmic star formation,
re-ionization of the of the intergalactic medium, and metallicity
histories of the universe.
\\ \\
The $Swift$ Gamma-Ray Burst Mission \cite{2004ApJ...611.1005G}
opens a window to explore the high redshift universe using GRBs.
Currently, there are only five GRBs with
spectroscopically-confirmed bursts with $z$ $>$ 5. The highest
redshift GRB detected thus far is GRB 080913 with redshift $\sim$
6.7 \cite{2008GCN..8225....1F}. The next highest redshift burst is
GRB 050904 which has redshift of $\sim$ 6.3
\cite{2006Natur.440..184K}. The remaining three are GRB 060927
with $z$ = 5.47 \cite{2007ApJ...669....1R}, GRB 050814 with $z$ =
5.3 \cite{2006AIPC..836..552J} and GRB 060522 with $z$ = 5.11
\cite{2006GCN..5155....1C}. All these high-z GRBs have been
discovered by the $Swift$ mission. Even though the number of
high-z GRBs are a handful, the Burst Alert Telescope (BAT;
\cite{2005SSRv..120..143B}), the primary instrument on board
$Swift$, has enough sensitivity to detect more high-z GRBs.
\\ \\
At present, the most difficult aspect of obtaining a
spectroscopically-measured redshift for a high-z GRB is to observe
the same object over the infrared band within a day after the
burst is detected by a space telescope. Useful infrared spectra of
GRBs require a large telescope, but the large telescopes in the
world have very limited available observing time and as such are
not in a position to do follow-up measurements on \emph{all} GRB
triggers. In this context, it would be useful to provide alerts to
ground-based observers when possible high-z GRBs are detected.
Previously, we have presented criteria \cite{2008AIPC.1000..166U}
which utilize trends with redshift to predict high-z GRBs with
some confidence. Now, with a larger sample some of those trends
have disappeared, and predicting high-z bursts requires
utilization of more observable parameters than before. In this
paper, we present revised criteria that enable the selection of
possible high-z GRBs using the prompt emission data from BAT.
\section{SCREENING HIGH-Z BURSTS USING BAT DATA}
In order to be effective, any high-z criteria need to be fast,
automated and reliable. We should also be able to make a decision
based on one or two orbits worth of data, which constrains the
number of available observable properties. We have looked at six
GRB BAT prompt emission properties to check for potential high-z
screening criteria. These properties are:
\begin{enumerate}
    \item \textbf{Peak Photon Flux:} Peak photon flux of the GRB
    measured in photons cm$^{-2}$ s$^{-1}$.
    \item \textbf{T90/Peak Photon Flux:} T90 is the time needed to
    accumulate from 5\% to 95\% of the counts in the 15--150 keV
    band. We normalize T90 by the peak photon flux to compensate
    for any detector threshold effects of BAT.
    \item \textbf{FFT Cutoff Frequency:} Cutoff frequency of the power spectrum when fitted
    by a broken power law curve.
    \item \textbf{Tau 50:} Sum of all burst time durations for which
    emission is greater that 50 \% of the total fluence.
    \item \textbf{Photon Index:} Photon index of the time integrated
    spectrum when fitted by a simple power law.
    \item \textbf{1-s Peak Photon Index:} Photon index of the peak 1 sec
    spectrum when fitted with a simple power law.
\end{enumerate}
We have used BAT event-by-event data to derive all the observable
parameters.
%To get the power spectrum we used ``IDL extract''
%software.
The redshift measurements were taken from online archives of the
Gamma-Ray Burst Online Index
(GRBOX\footnote{http://lyra.berkeley.edu/grbox/grbox.php}) and
verified them using the GCN
circulars.\footnote{http://gcn.gsfc.nasa.gov}
\\ \\
%We expect to see a number of distinct properties in the prompt
%emission for a high-z GRB. Since high-z GRBs have occurred at a
%larger luminosity distance than typical z=1 GRBs, we would expect
%them to have a lower peak photon flux. Moreover, due to cosmic
%time dilation, the light curve of these high-z GRBs should have
%longer durations and smaller FFT cutoff frequencies. Furthermore,
%the prompt emission spectrum of the \emph{Swift} high-z bursts is
%not soft.
%\\ \\
We might expect to see a number of distinct properties in prompt
emission from high-z GRBs in comparison to those with low-z. Being
at large redshift, high-z GRBs would have longer time durations
than similar GRBs with lower redshifts.  Contrary to this
expectation, we have found that some high-z GRBs have relatively
short duration. Also, because of the cosmic dilation effect,
high-z GRBs emission variabilities should have lower FFT cutoff
frequencies in comparison to the same type of GRB in the burst
rest frame.
\begin{figure}[htp]
  \includegraphics[height=.16\textheight]{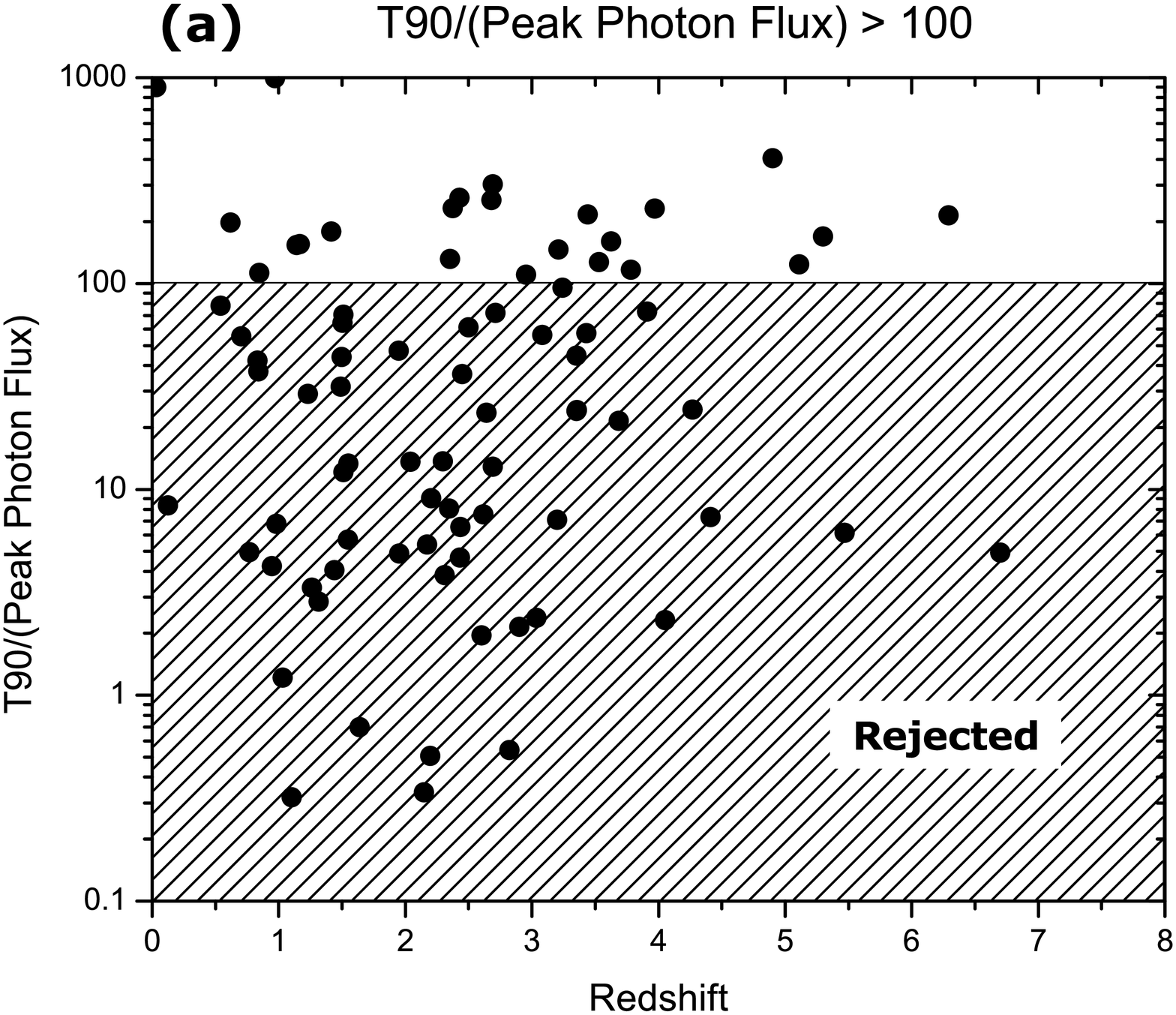}
  \includegraphics[height=.16\textheight]{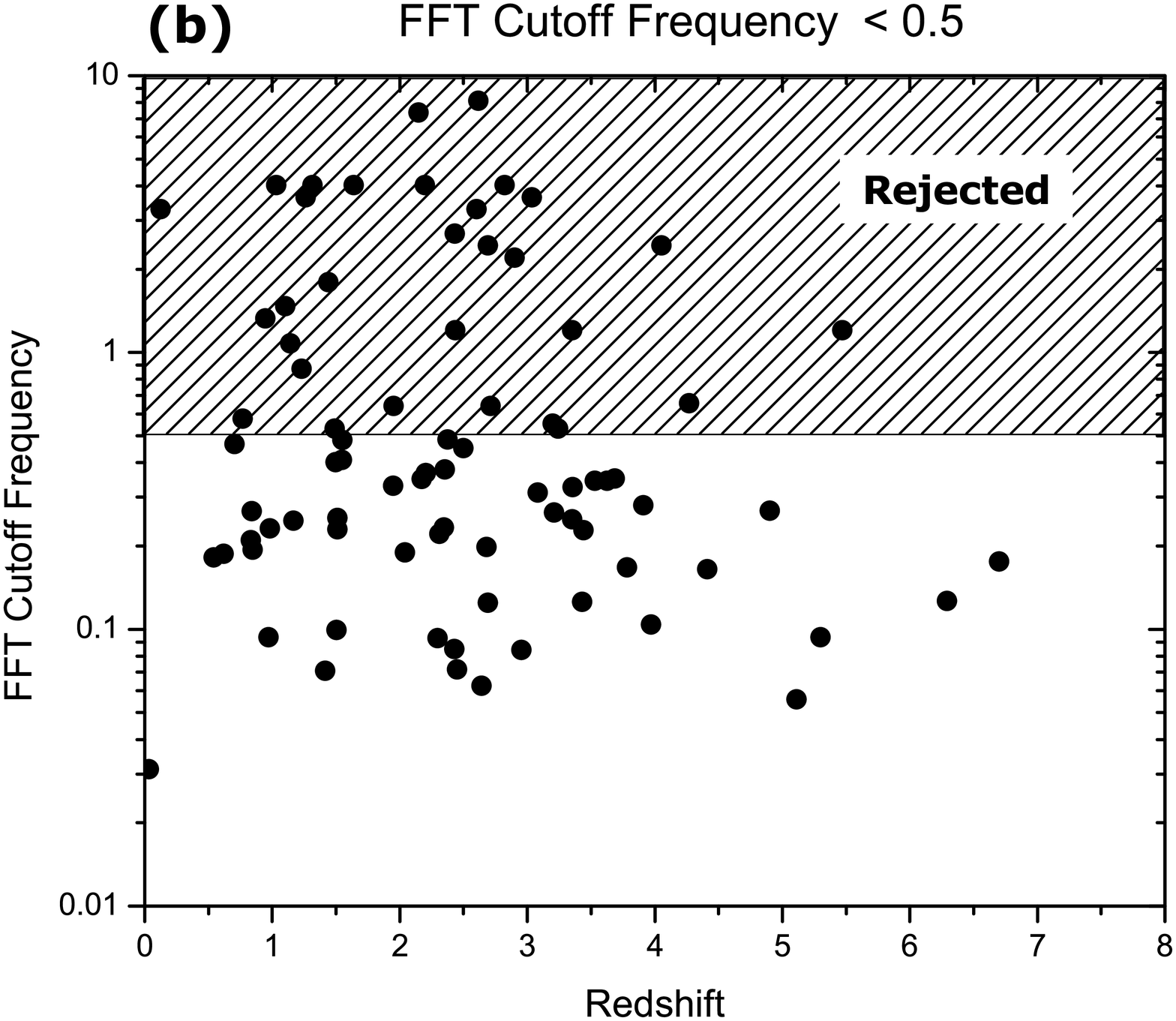}
  \includegraphics[height=.16\textheight]{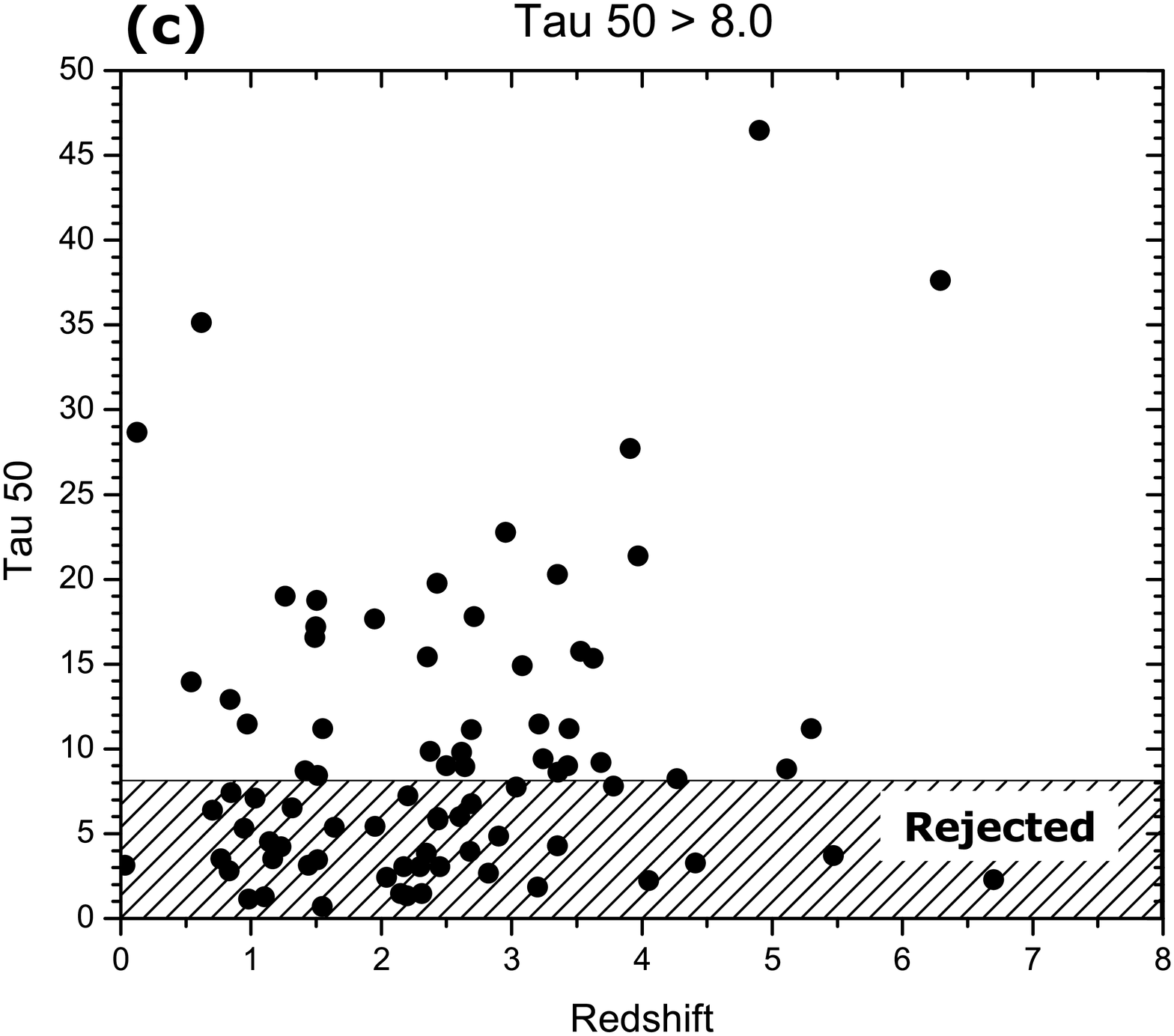}
\end{figure}
\begin{figure}[htp]
  \includegraphics[height=.16\textheight]{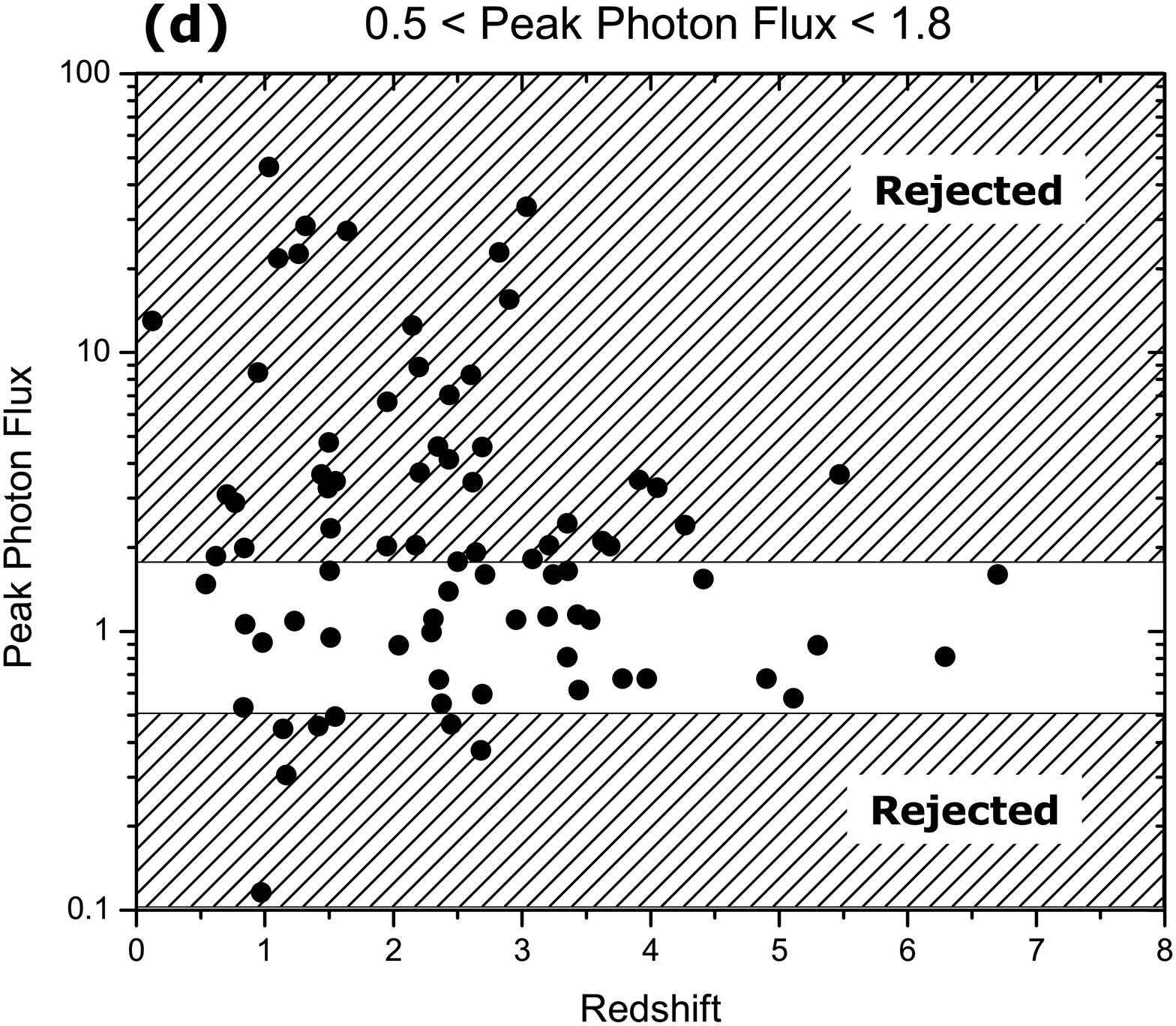}
  \includegraphics[height=.16\textheight]{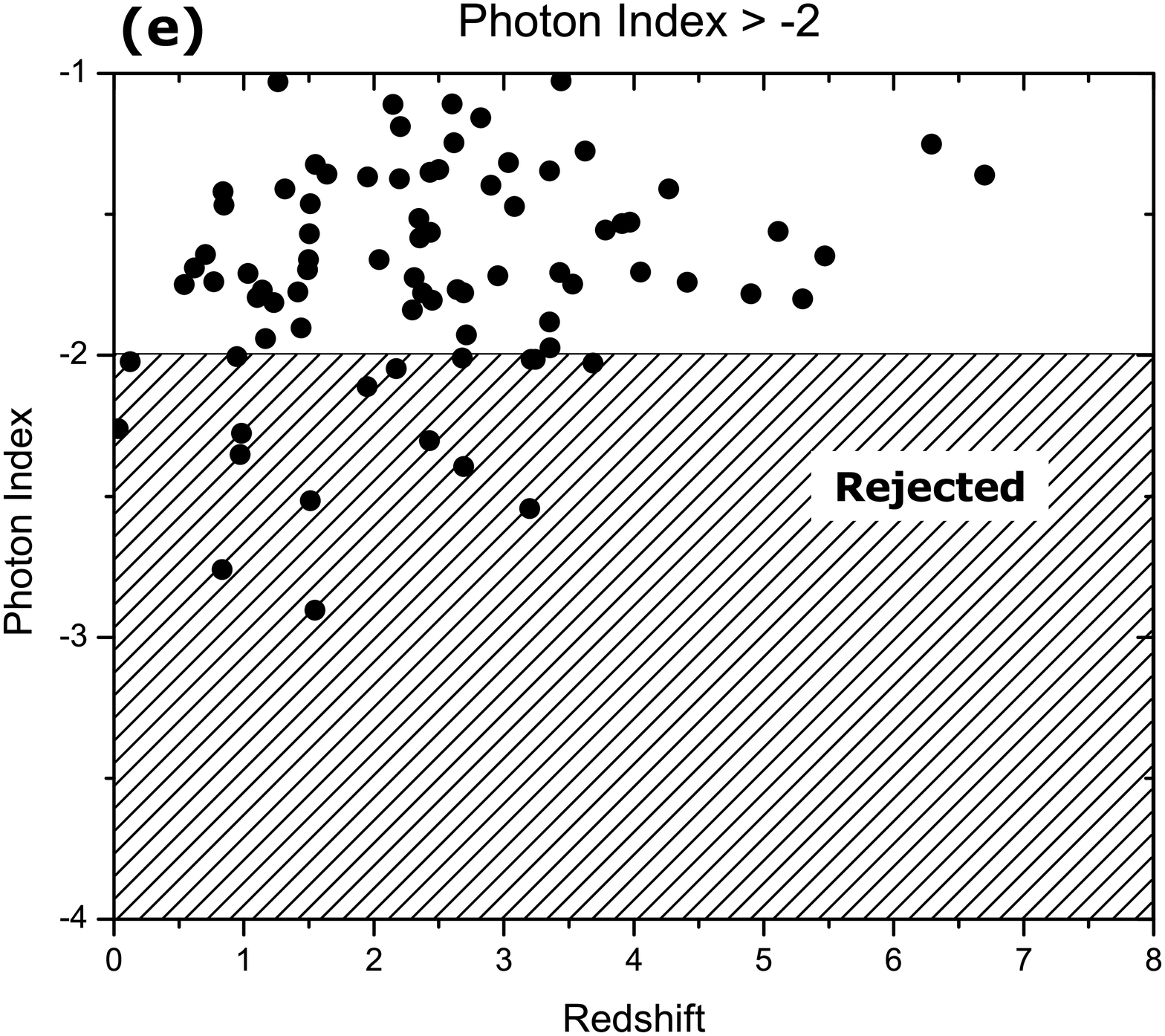}
  \caption{First Criteria; (a) T90/(Peak Photon Flux) $>$ 100, (b)
  FFT Cutoff Frequency $<$ 0.5, (c) Tau 50 $>$ 8.0, (d) 0.5 $<$
  Peak Photon Flux $<$ 1.8, and (e) Photon Index $>$ -2. This
  criterion is aimed at selecting high-z burst which has smooth
  long light curves.
  \label{Fig01}}
\end{figure}
Similarly, the Doppler effect would soften the detected emission
for a given type of GRB at high z.  However, the prompt spectrum
of the Swift high-z bursts is relatively hard. Moreover, high-z
GRBs occur at a larger luminosity distance than typical $z=1$
GRBs, which results in a lower peak photon flux. These
observations may arise from two or more distinct populations of
high-z GRBs. This led us to impose two distinct high-z criteria
instead of one.
\\ \\
%We expect to see a number of distinct properties in the prompt
%emission for a high-z GRB. Since high-z GRBs occur at high
%redshifts, we would expect them to have longer time durations, due
%to cosmic time dilation. Contrary to this expectation we have
%found that some high-z GRBs have relatively shorter durations
%which may result from two different populations of high-z GRBs.
%\begin{figure}[htp]
%  \includegraphics[height=.16\textheight]{z_vs_T90PPF1}
%  \includegraphics[height=.16\textheight]{z_vs_FFT_Freq1}
%  \includegraphics[height=.16\textheight]{z_vs_Tau50_1}
%\end{figure}
%\begin{figure}[htp]
%  \includegraphics[height=.16\textheight]{z_vs_PPF1}
%  \includegraphics[height=.16\textheight]{z_vs_PI_1}
%  \caption{First Criteria; (a) T90/(Peak Photon Flux) $>$ 100, (b)
%  FFT Cutoff Frequency $<$ 0.5, (c) Tau 50 $>$ 8.0, (d) 0.5 $<$
%  Peak Photon Flux $<$ 1.8, and (e) Photon Index $>$ -2. This
%  criterion is aimed at selecting high-z burst which has smooth
%  long light curves.
%  \label{Fig01}}
%\end{figure}
%This prompted us to consider two distinct high-z criteria instead
%of one. Also due to the same cosmic dilation effect, we expect
%high-z GRBs to have smaller FFT cutoff frequencies, assuming
%variability at the burst rest frame is more or less the same.
%Moreover, high-z GRBs occur at a larger luminosity distance than
%typical $z=1$ GRBs, which results in a lower peak photon flux.
%Furthermore, due to cosmological Doppler effect we expect high-z
%burst to be soft, however, again contrary to our expectation, the
%prompt emission spectrum of the \emph{Swift} high-z bursts is
%hard.
%\\ \\
Our first high-z criteria, which is aimed at catching long, smooth
high-z GRBs, is a set of five cuts as shown in Fig.~\ref{Fig01}.
Our second high-z criteria is a set of four cuts as shown in
Fig.~\ref{Fig02}. With the limited available sample of GRBs having
known redshifts, the probability of a GRB being selected by both
criteria with a $z$ $>$ 5.0 is 24 \%, with a $z$ $>$ 4.0 is 38 \%
and with a $z$ $>$ 3.5 is 48 \% (Table~\ref{tab:01}). This is
roughly factor of four improvement over not having any high-z
criteria. Based on an analysis of Swift existing burst data, these
criteria imply an average alert rate of $\sim$ one per 1.5 months.
In addition to the above criteria we will use excess NH
measurements from XRT\footnote{X-ray Instrument onboard $Swift$}
to do a secondary check before issuing an alert
~\cite{2007AJ....133.2216G}.
\begin{figure}[htp]
  \includegraphics[height=.16\textheight]{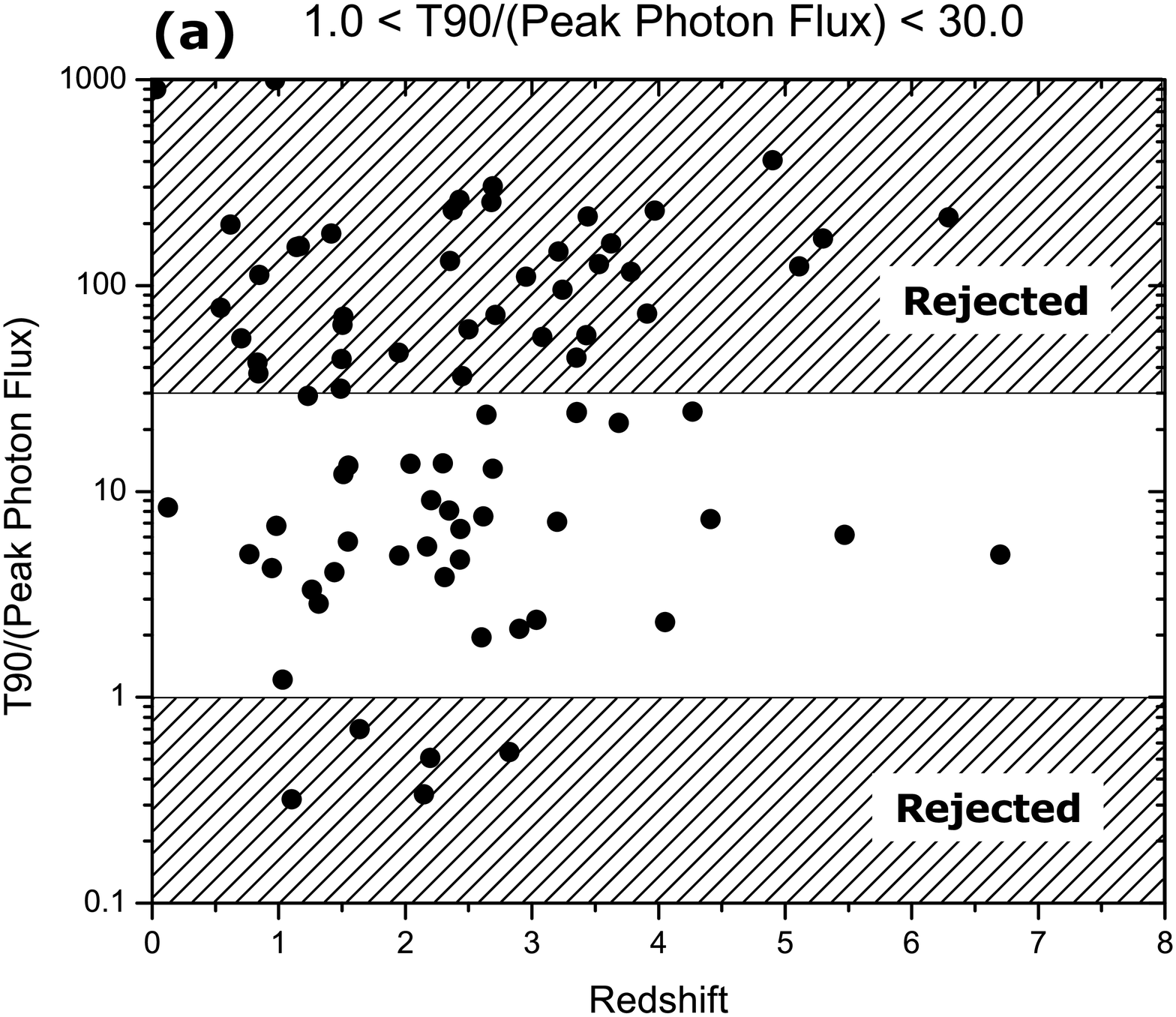}
  \includegraphics[height=.16\textheight]{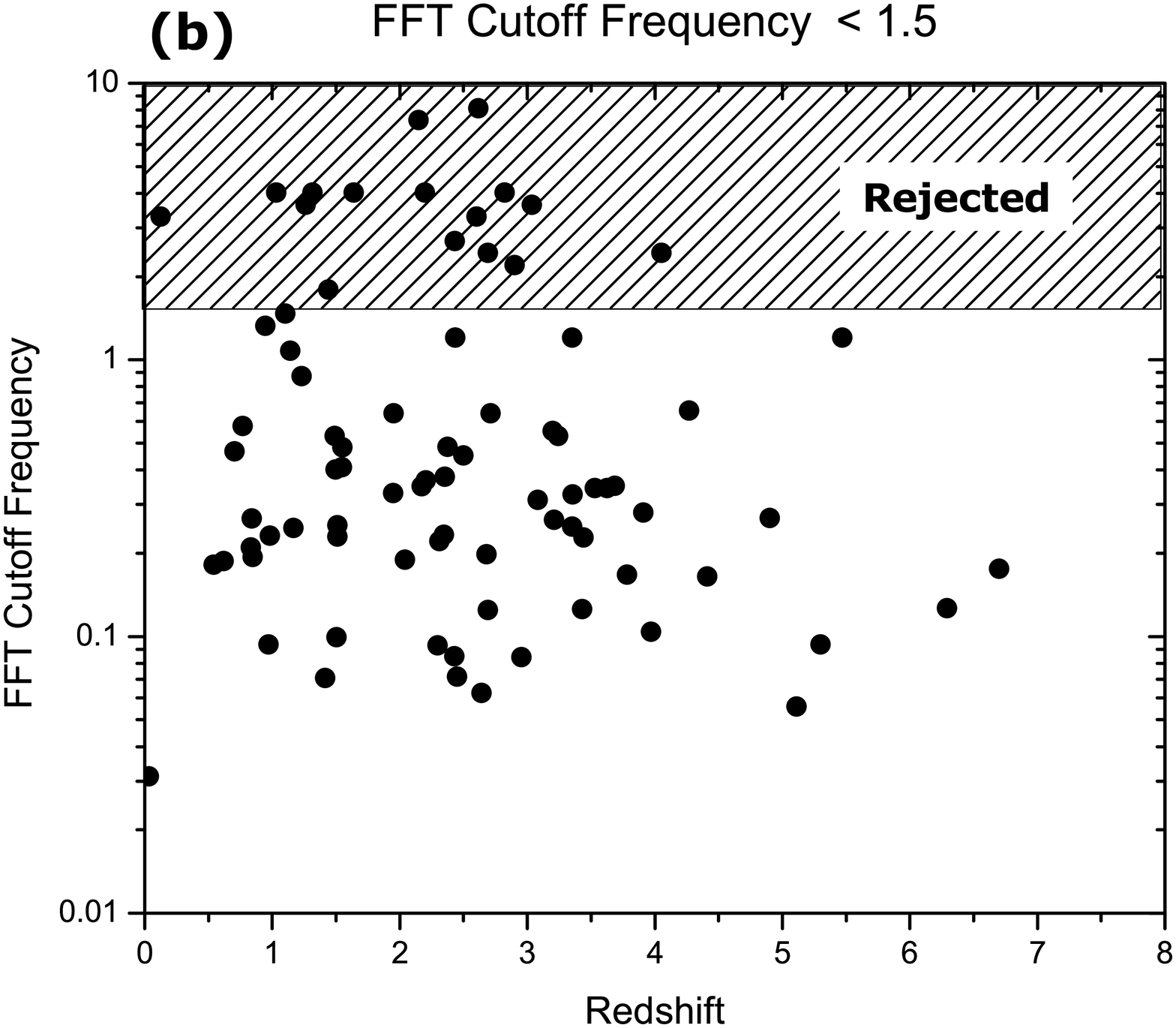}
  \includegraphics[height=.16\textheight]{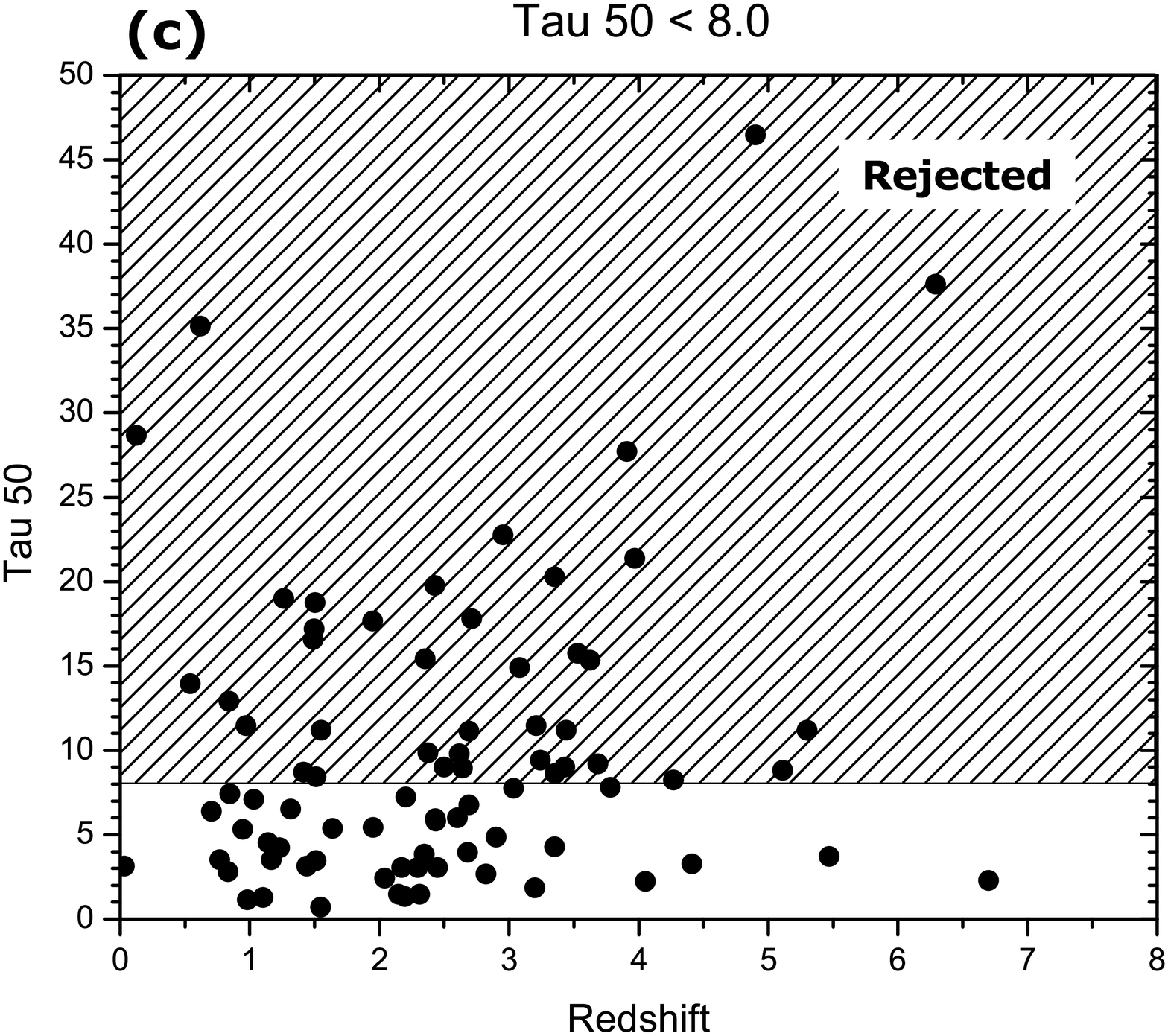}
  \includegraphics[height=.16\textheight]{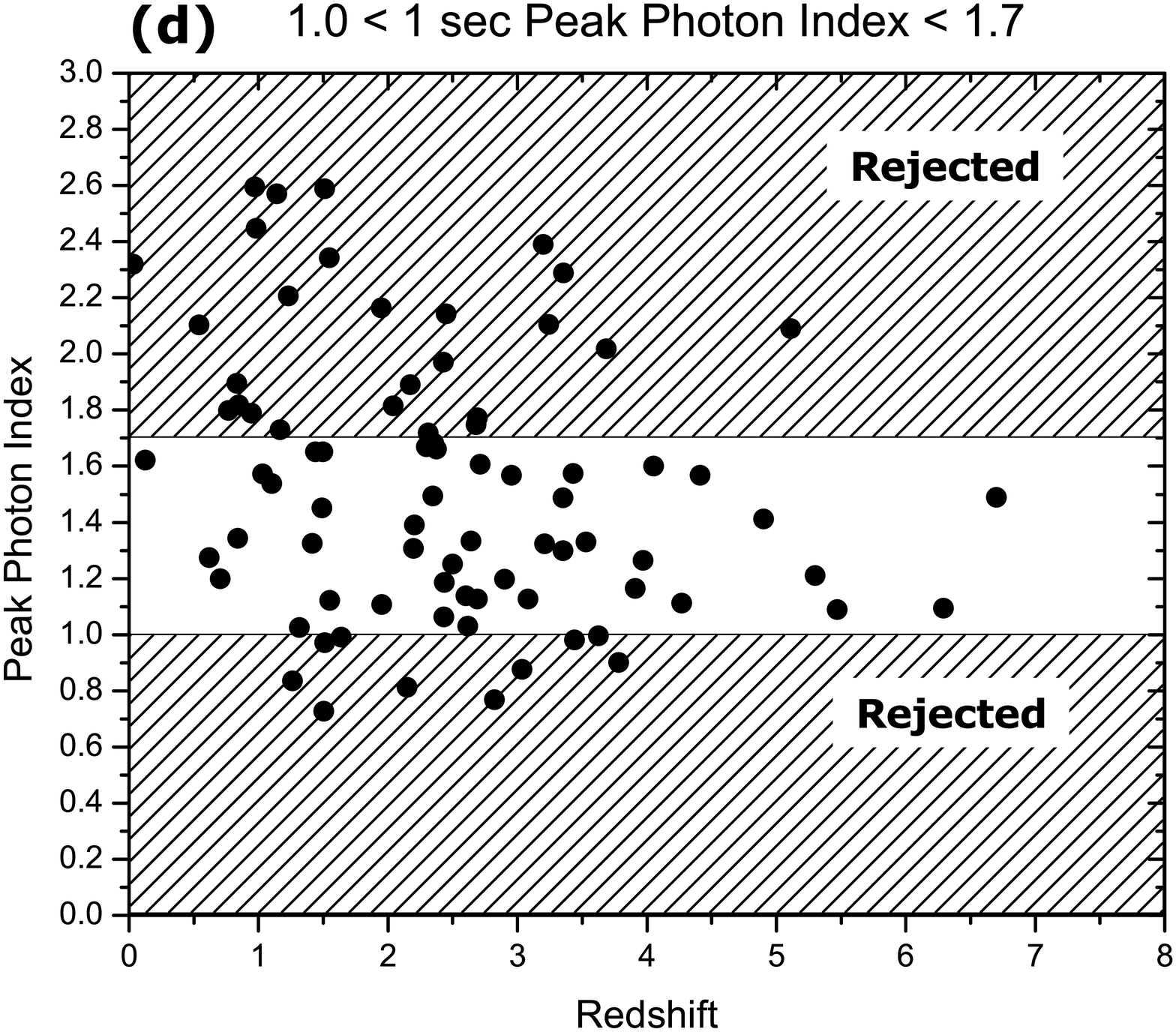}
  \caption{Second Criteria; (a) 1.0 $<$ T90/(Peak Photon Flux) $<$ 30.0, (b)
  FFT Cutoff Frequency $<$ 1.5, (c) Tau 50 $<$ 8.0, and (d) 1.0 $<$
  1 sec Peak Photon Index $<$ 1.7. This criterion is aimed at selecting high-z burst which has
  short light curves.
  \label{Fig02}}
\end{figure}
\begin{table}
\begin{tabular}{lrrrr}
\hline
    $$
  & Criteria 01
  & Criteria 02
  & No Criteria
  & Both Criteria \\
\hline
Redshift $\ge$ 5.0 & 30 \% & 18 \% & 6 \% & 24 \%\\
Redshift $\ge$ 4.0 & 40 \% & 36 \% & 11 \% & 38 \%\\
Redshift $\ge$ 3.5 & 60 \% & 36 \% & 18 \% & 48 \%\\
\hline
\end{tabular}
\caption{Comparison of success probability for high-z criteria and
no high-z criteria.} \label{tab:01}
\end{table}
%\begin{table}
%\begin{tabular}{lrlr}
%\hline
%    \tablehead{1}{l}{b}{GRB Name (Criteria 01) }
%  & \tablehead{1}{r}{b}{Measured Red-shift}
%  & \tablehead{1}{l}{b}{GRB Name (Criteria 02)}
%  & \tablehead{1}{r}{b}{Measured Red-shift}  \\
%\hline
%GRB 050730 & 3.97 & GRB 050505 & 4.27\\
%GRB 050814 & 5.30 & GRB 050908 & 3.35\\
%GRB 050904 & 6.29 & GRB 051109A & 2.35\\
%GRB 060115 & 3.53 & GRB 060124 & 2.30\\
%GRB 060510B & 4.90  & GRB 060223A & 4.41\\
%GRB 060522 & 5.11 & GRB 060927 & 5.47\\
%GRB 061110B & 3.44 & GRB 080210 & 2.64\\
%GRB 070110 & 2.35 & GRB 080319C & 1.95\\
%GRB 070411 & 2.95 & GRB 080413A & 2.43\\
%GRB 080905B & 2.37 & GRB 080804 & 2.20\\
%- & - & GRB 080913 & 6.70\\
%\hline
%\end{tabular}
%\caption{Selected GRBs using the two high-z criteria from Swift
%GRBs with known red-shifts.} \label{tab:02}
%\end{table}
\section{Conclusion}

We have investigated the possibility of screening high-z GRBs
based on BAT prompt emission data. Our analysis seems to indicate
there are two populations of high-z GRBs; hence we have two sets
of criteria. The first criteria tries to filter out long, smooth
high-z burst while the second criteria aims at shorter duration
high-z bursts. Most of the high-z bursts seems to be weak and
hard. There is roughly factor of four improvement over not having
any high-z criteria at all. Screening high-z GRBs based only on
BAT prompt emission data remains very challenging and work is
continuing to search for more reliable BAT high-z indicators.

\end{document}